# Coherent microwave radiation emitted during the process of stimulated Raman scattering in weakly compressed hydrogen: experimental studies, together with an attempt to interpret this microwave emission

--------------------

**B. Oksengorn**

Laboratoire d'Ingénierie des Matériaux et des Hautes Pressions

CNRS, Université Paris 13

99 av. Jean-Baptiste Clément, F-93430 Villetaneuse, France

telephone: 33 (0)1 49 40 34 46

Fax : 33 (0)1 49 40 34 14

E-mail: boris.oksengorn @ limhp.univ-paris13.fr

**Abstract** : A coherent microwave radiation had been observed, in the first time to our knowledge, during experiments of stimulated Raman scattering (SRS) in weakly compressed hydrogen, simultaneously with the emission of Raman lines. The intensity of this microwave radiation had been studied as a function of the gas pressure, and the product of the laser and Stokes wave powers. Moreover, the intensity spread from the propagation axis of the laser beam had been measured, just as the wavelength of centimetric order of magnitude, and an estimate of the peak power for this microwave emission.

A phenomenological interpretation of this coherent microwave radiation in SRS with hydrogen is proposed, assuming that an intermediate quasiparticle state, vibrationally excited during a very short time ($10^{-13}$–$10^{-12}$ s), should be taken into account, just as a macroscopic longitudinal electric field. Moreover, a nonlinear coupling of anharmonic character, between the Stokes wave and the longitudinal waves associated to the quasiparticle system, is taken into account. Besides, this system might get a "piezoelectric" character under the mechanical action of the laser field. On the other hand, a Cherenkov-type and an anomalous relativistic Doppler effects are also assumed. Then, it has been found that the in phase quadrature part of the longitudinal component of the polarisation associated to the quasiparticle waves, might be the source of the coherent microwave radiation, emitted during parametric instabilities, the duration of which should be $10^{-13}$-$10^{-12}$ s.

Concerning the SRS process itself, we note that an enhancement of the Raman gain might be due to an additional term arising from the electro-optic effect induced by the electric field of the laser beam; this term should be effective, only for durations of the same order of magnitude as for the microwave emission.

-----------------

--------------

**I-Introduction.-** First of all, we want to present some qualitative results, obtained a long time ago, when studies of radiations emitted in the infrared region, during SRS experiments performed with a Q-switched ruby laser in weakly compressed hydrogen, had been carried out. As a matter of fact, using a filter of silicon after the pressure cell, we had observed fortuitously a new kind of radiation provided by our infrared detector, a high-speed Ge:Au one working at the liquid nitrogen temperature and sensitive to infrared radiations up to 9 µm, and supplied by the manufacturer Santa Barbara Research Centre, Goleta, Cal.(USA), as we are going to show it below. It is truthful that possibilities of common artefacts had existed, due to the radiations emitted by the flash-lamp of our Q-switched ruby laser, and by the electric discharges in the laser equipment, or by the plasma induced in hydrogen by the laser field. Consequently, different tests were carried out to show that we had observed effectively a new type of radiation, as follows: 1) the signal obtained by means of our Ge:Au detector had a shape which is very different from those of the laser, Raman and flash-lamp pulses (see Fig.2); 2) at the pressure of one bar no stimulated Raman lines had been observed, and no signal detected with our Ge:Au detector; 3) with argon gas instead of hydrogen, no signal had be seen, originating from the light emitted by the laser-induced plasma; 4) by means of an aperture of 3 mm diameter laid on the Ge:Au detector, only infrared radiations due to SRS in weakly compressed hydrogen had been observed; 5) with a black paper before the normal aperture of 2 cm diameter, only the new radiation had been observed; 6) in liquid water and gaseous ammonia, absorptions of this radiation were detected.

Afterwards, new experiments have been carried out more recently, to yield quantitative data as a function of several parameters, published in a short paper [1]. Nevertheless, a more detailed description of these experimental studies is given in Section II of the present paper. Furthermore, in Section III is presented an attempt to interpret this new type of radiation, appearing together with SRS in hydrogen. In Section IV a large discussion is given about similarities and differences between the physical process studied in this paper and others nonlinear phenomena. Finally, Section V is devoted to some remarks about the experimental SRS gains, which had been found above the SRS threshold, very much greater than the theoretical ones [2].

**II-Experimental set-up and results.-**For our new experiments a Nd:Yag pulsed laser has been used, the oscillator of which was Q-switched by a Pockels cell, followed by an amplifier. The laser beam was linearly polarised, monomode transverse, and almost longitudinally monomode, with 2 or 3 modes only, by using an intracavity FP glass disk. The laser pulse had a spectral width of 0.08 cm$^{-1}$, and its duration was equal to 15 ns. Besides, the laser power at $\lambda=1.064$ µm was equal to a maximum of 30 MW, reduced to 10 MW at $\lambda=532$ nm by frequency-doubling, and the laser working was effected at the repetition rate of 10Hz..

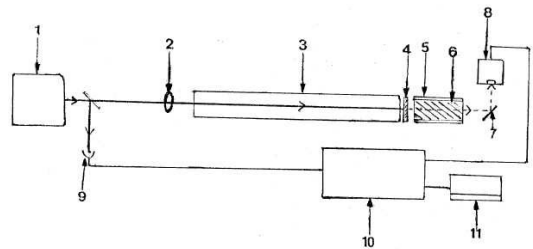

Fig.1-Experimental set-up : 1=Nd-Yag laser ; 2=focusing lens; 3=high pressure cell;4=silicon disk of 5 mm thickness used as filter; 5=Cu tube; 6=polyethylene rod of 20 cm length; 7=Cu mirror; 8=Ge:Au detector; 9=light detector for triggering; 10=pulse analyser; 11=computer.

The experimental set-up is described on Fig.1: the laser beam was focused by means of a lens of 50 cm focal length in the high-pressure optical cell of 1m length and 2 cm inner diameter. At the end of the cell two filters were put to stop all visible and infrared radiations: one was a silicon disk of 5 mm thickness, the other a polyethylene rod of 20 cm length and of 2 cm diameter, which was put in a Cu tube. The efficiency of this filter system had been checked by means of a InSb detector, cooled at liquid nitrogen temperature, the sensitivity of which was 100 times higher than that of the Ge:Au detector in the infrared region of 5 µm: no signal was detected during SRS in hydrogen. Behind the filter system a Cu mirror was set for reflecting the microwave radiation to the Ge:Au detector. Finally, data were collected by means of a pulse analyser or a digital oscilloscope.

On Fig.2 is displayed the signal observed with the Ge:Au detector during a SRS experiment; it is strongly asymmetrical, with a short rise time estimated to 7 ns, the Stokes pulse duration being equal to 10 ns; the exponential decay of this signal has a characteristic time of 65 ns. Moreover, in the inset is shown a signal which is very similar to the previous one, and observed in experiments carried out by Manenkov et al., concerning thermal breakdown of excitons by a microwave field; these particles were produced previously by laser excitation of free carriers in a pure germanium disk cooled at 1.3 K, followed by a binding process [3]. In addition, Keldysh et al. had improved their experiments, and had obtained for the rise and fall-off times of the breakdown pulses the values of 100 and 200 ns, respectively [4]. Accordingly, we think that in our SRS experiments with hydrogen a coherent emission of microwave pulses was also detected, because the sensitivity of our detector to microwave radiation might be due to the fact that under the influence of the polarisation field, induced by a dc 40 V voltage, an equilibrium between free carriers, excitons and / or exciton complexes might appear in the Ge:Au disk, cooled at 77 K. After that, the heating of free carriers by the assumed microwave radiation would be the cause of the exciton breakdown, giving rise to the increasing part of the signal, whereas its decreasing part should correspond to the characteristic time of exciton recombination.

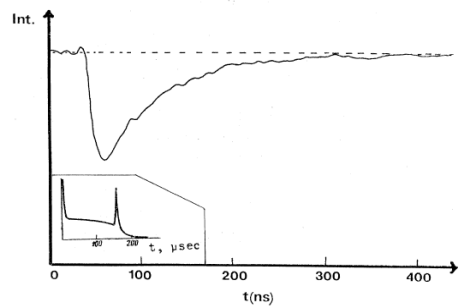

Fig.2- Profile of the pulse signal obtained by means of the Ge:Au detector, and due to the absorption of microwave radiation. Inset: the same obtained with pure germanium in ref. [3] (see the text).

Besides, experiments had been performed to study the microwave intensity as a function of the gas pressure. On Fig.3 is shown this dependence, in the case of a SRS experiment with excitation by the 1.064 μm laser beam, while in the inset this dependence is depicted for a shorter region of the gas pressure, when the SRS experiment was performed by means of a 532 nm laser excitation. It can be seen that the microwave intensity is strongly increasing in the very low pressure range. Afterwards, a maximum appears in the 8-10 bar region, followed by an exponential decreasing up to about 20 bar; at higher pressure the signal intensity is nearly constant. Moreover, the exponential behaviour of the decreasing part is confirmed on Fig.4, where the neperian logarithm of the signal intensity follows a linear variation with the number density of hydrogen.

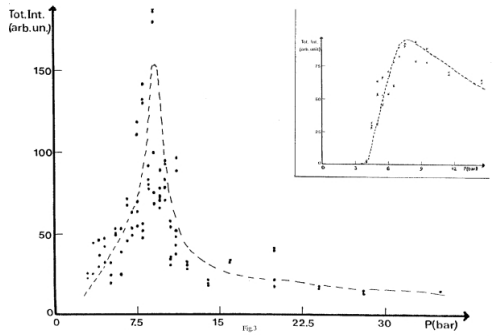
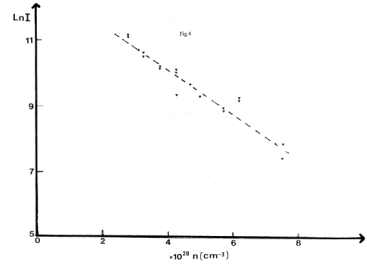

Fig.3-Variation of the microwave intensity as a function of hydrogen pressure exciting laser at λ=1.064 μm). Inset: the same with exciting laser at λ=532 nm. The dashed lines are only for the eye.

Fig.4- Linear variation of the neperian logarithm of the microwave intensity, versus the hydrogen number density. The dashed line is only for the eye.

In addition, the microwave intensity, measured at the pressure of about 10 bar and with the 532 nm laser excitation of the signal, follows a linear variation with the product of the laser and Stokes beam powers, as shown on Fig.5. This linear relationship should point out that the process of coherent microwave emission in hydrogen should be due to a nonlinear optical phenomenon, like a parametric amplification.

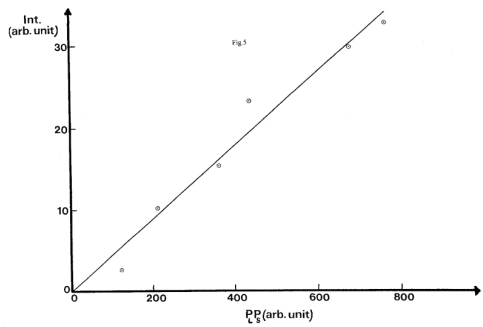
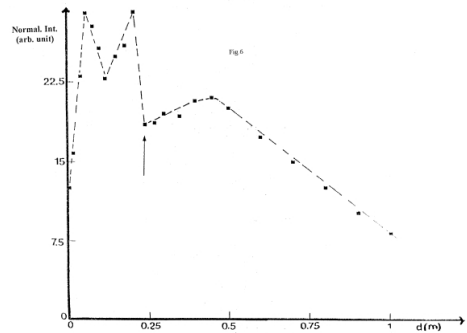

Fig.5- Linear variation of the microwave intensity, as a function of the product of the laser and Stokes beam powers.

Fig.6- Evolution of the microwave normalised intensity, versus the distance between detector and propagation axis of the laser beam. The arrow might point to the limit of the Cherenkov cone. The dashed line is only for the eye.

Moreover, on Fig.6 the evolution of the normalized intensity of the microwave signal is shown as a function of the distance between the detector, set at 1m from the end of the high pressure cell, and the propagation axis of the laser beam. One can see that the signal may be detected during a large displacement of the detector, and that its profile is modulated, probably by diffraction effects.

On the other hand, we had determined the wavelength of the microwave radiation by using a home-made wavemeter, with which a wavelength of about 1.4 cm had been measured, corresponding to a frequency of 20 GHz. Moreover, we had attempted to determine the microwave power by means of a detector manufactured by Spacek Labs. Inc.

(model DK-2); the sensivity of this detector was 2.5 V/mW in the K band (18 to 26 GHz). An average of 6 nW had been obtained, which corresponds for the peak power to an order of magnitude of 10 µW. However, this power is not the true value emitted during the SRS process, because different effects are not taken into account during the propagation of the microwave emission, as reflections, absorptions, diffractions and scatterings.

Finally, during a recent SRS experiment with hydrogen, we have been able to detect a backscattered microwave radiation, the maximum intensity of which has been found to be at the same pressure of about 10 bar, as in the case of forward scattering.

**III-Theory.-**First of all, we propose for the slope $v_o$ of the linear relationship depicted on Fig.4, the value of which is equal to $6.62 \cdot 10^{-21}$ cm$^3$, the following physical meaning: this parameter should correspond to the volume of a quasiparticle, being composed of a characteristic number of molecules vibrationally excited during the SRS process. An order of magnitude of this number can be calculated by means of the model used by Hirschfelder et al. for a gas, the face-centered cubic lattice gas [5]: at low density the free volume occupied by one molecule is given by $v=3.18 \sigma^3$, where $\sigma$ is the hard sphere diameter of the $H_2$ molecule, equal to 0.287 nm in the case of a Lennard-Jones potential. By comparing the volume $v$, equal to $7.52 \cdot 10^{-23}$ cm$^3$ to the volume $v_o$ of the quasiparticle, an order of magnitude of about one hundred of vibrationally excited oscillators is found to make up one quasiparticle. Our hypothesis might be corroborated by the following points: owing to the excitation by the coherent laser beam, the vibrationally excited oscillators are all in the same phase, and may interact strongly between them, through their anharmonicity. Moreover, if the density of these interacting particles becomes appreciable, what is the case for the Stokes wave intensity above the SRS threshold in the present instance, a cooperative effect should be present, leading to a kind of "localisation" for the excited oscillators, according to the model developed by Kindall et al. for the case of anharmonic solids [6]. In other respects, let us note that Boling et al. had used a model based on the anharmonic oscillator to predict nonlinear refractive index changes in optical solids, and had obtained a rather good agreement between experimental data and their empirical relationships [7].

On the other hand, we assume that the quasiparticle system is an intermediate state which should have a very short lifetime $\tau_R$, of about $10^{-13}$-$10^{-12}$s, compared to the laser pulse duration. Besides, we admit that the main interaction between these quasiparticles is of coulombian type, yielding a macroscopic electric field, longitudinally polarised. Moreover, it is assumed that the stationary wave associated to the quasiparticle system is modulated at the angular frequency $\omega_R$ of the quasiparticle vibration, which should be very close to the angular frequency $\omega_v$ of the isolated molecular vibration, but slightly higher than the latter.

Furthermore, another important hypothesis should be assumed: the quasiparticles are distorted by the electrostriction effect due to the laser field, what gives infrared active their vibrational mode, together with the advent of an induced dipole moment and an anomalous dispersion of the refractive index at the frequency $\omega_R$ ; another effect of this distortion leads to the lack of the centre of inversion for the quasiparticle system, which gets a "piezoelectric" character [8].

Afterwards, we will study the interaction between the dielectric medium and the laser light, by assuming that the wave associated to the quasiparticle system, and being stationary compared to the velocity of light in the medium, may be split into two progressive waves, propagating the one to forward, the other to backward, with a phase velocity very close to the light velocity in vacuum c, but higher than the light velocity in the dielectric medium. Thus, the stimulated processes, like Stokes and microwave emissions, should be distinguished by a Cherenkov-type effect [9].

We assume also that the "piezoelectric" character of the quasiparticle system can be depicted by two parameters, induced by the laser field: one is $C_{ind.}^R$, which denotes the atomic displacement, whereas the other is $C_{ind.}^E$, which is an electro-optic coefficient. These parameters are tensors of third rank, and can give rise to a nonlinear coupling of anharmonic origin, between light and matter waves, as shown by Merten et al. in the case of piezoelectric cubic solids [10-13]. In our case, we propose that the coupled mode photon-quasiparticle might be so-called "anharmonic longitudinal polariton" [12]. Furthermore, taking into account only the coupling with the Stokes wave in the parametric approximation, and writing: 1) $R_z$ for the longitudinal component of the normal coordinate **R** of a quasiparticle, the z axis being the laser and Stokes wave propagation axis; 2) $P_z$ for the longitudinal component of the polarisation density, associated to the quasiparticle waves; 3) $(P_x^S)^*$ for the transverse component of the complex conjugate of the polarisation density associated to the Stokes wave, the x axis being the direction of polarisation for the laser and Stokes waves, we are able to obtain the three following equations, according to ref. [10-13]:

$$-\omega_R^2 R_z = BE_z + C_{ind.}^R E_x^L (E_x^S)^* \quad (1a)$$

$$P_z = [(n_R^2 - 1)/4\pi] E_z + C_{ind.}^E E_x^L (E_x^S)^* \quad (1b)$$

$$(P_x^S)^* = [(n_L^2 - 1)/4\pi](E_x^S)^* + C_{ind.}^E E_x^L E_z + C_{ind.}^R R_z \quad (1c)$$

In these equations the wave damping and the dispersion of the refractive index are neglected, excepting the anomalous dispersion at the frequency $\omega_R$; the parameter B is the coefficient of the mechanical-electric linear coupling, and is a scalar which we assume to be a function of the anharmonicity of the quasiparticle vibration; $E_z$ is the longitudinal electric field, $E_x^L$ and $E_x^S$ are the electric fields of the laser and Stokes waves, respectively; finally, $n_L$ and $n_R$ are the refractive indices at the laser frequency and the frequency $\omega_R$, respectively.

Afterwards, it is necessary to introduce the dielectric displacements $D_z$ and $(D_x^S)^*$ instead of $P_z$ and $(P_x^S)^*$, given by

$$D_z = \varepsilon_z^{(1)} E_z + \varepsilon_{zx}^{(2)} (E_x^S)^* \quad (2a)$$

$$(D_x^S)^* = \varepsilon_{zx}^{(2)} E_z + \varepsilon_{zx}^{(3)} (E_x^S)^* \quad (2b)$$

where the dielectric functions $\varepsilon_z^{(1)}$, $\varepsilon_{zx}^{(2)}$ and $\varepsilon_{zx}^{(3)}$ are functions of the frequency $\omega_R$, as follows

$$\varepsilon_z^{(1)} = n_R^2 - 4\pi B^2 / \omega_R^2$$

$$\varepsilon_{zx}^{(2)} = 4\pi E_x^L (C_{ind.}^E - B C_{ind.}^R / \omega_R^2)$$

$$\varepsilon_{zx}^{(3)} = n_L^2 - [4\pi (C_{ind.}^R)^2 / \omega_R^2](E_x^L)^2$$

Besides, $D_z$ and $(D_x^S)^*$ must fulfil Maxwell's equations, which lead to the following relations

$$-\varepsilon_z^{(1)} E_z - \varepsilon_{zx}^{(2)} (E_x^S)^* = 0 \quad (3a)$$

$$-\varepsilon_{zx}^{(2)} E_z + (n_L^2 - \varepsilon_{zx}^{(3)})(E_x^S)^* = 0 \quad (3b)$$

The solution for this system of two equations exists, only if the determinant of its coefficients vanishes, leading to the relation

$$\varepsilon_z^{(1)} (n_L^2 - \varepsilon_{zx}^{(3)}) + (\varepsilon_{zx}^{(2)})^2 = 0 \quad (4)$$

From the solution of this equation (4) it is possible to determine the ratio $C_{ind.}^R / C_{ind.}^E$ by assuming the inequality $4\pi B^2 / n_R^2 << \omega_R^2$; in that case one obtain the following relation

$$C_{ind.}^R / C_{ind.}^E = 4\pi B / n_R^2 \pm i\, 2\pi^{1/2}\, \omega_R / n_R \quad (5)$$

where it can be seen that these two parameters are complex quantities, while in the case of piezoelectric cubic solids, they are only real.

It is also possible to calculate the dispersion functions for the coupled waves by means of the effective dielectric function, given by

$$\varepsilon_{\text{eff}} = [\varepsilon_{zx}^{(3)} \varepsilon_z^{(1)} - (\varepsilon_{zx}^{(2)})^2]/\varepsilon_z^{(1)}$$

$$\varepsilon_{\text{eff}} = n_L^2 - (32\pi^2 n_R^{-2})|C_{\text{ind.}}^E|^2|E_x^L|^2$$

The results for the dispersion functions can be represented by complex-conjugate values of the wave vectors, as follows

$$K_1 = K_L - i\, n_L\,[1 - 16\pi^2\,(n_L^2\, n_R^2)^{-1}|C_{\text{ind.}}^E|^2|E_x^L|^2]\,(\omega_L - \omega_R)/c$$

$$K_2 = K_L + i\, n_L\,[1 - 16\pi^2\,(n_L^2\, n_R^2)^{-1}|C_{\text{ind.}}^E|^2|E_x^L|^2]\,(\omega_L - \omega_R)/c$$

According to ref.[10-13], the imaginary parts of these relations should correspond to an exponential amplification for the two progressive waves associated to the quasiparticle system, one to forward and the other to backward.

Moreover, from equations (3a) and (5) it is possible to obtain the following relation between the longitudinal electric field $E_z$, and the laser and Stokes fields

$$E_z = -4\pi\, n_R^{-2}\, C_{\text{ind.}}^E\, E_x^L (E_x^S)^* \pm i\, 8\pi^{3/2}\, n_R^{-3}\, B\omega_R^{-1}\, C_{\text{ind.}}^E\, E_x^L (E_x^S)^* \quad (6)$$

It is interesting to note that this longitudinal field induces a polarisation $P_z$, to which two progressive waves may be associated, with the same phase velocity as assumed above. On the other hand, it can be seen that a part of these longitudinal polarisation waves are in phase quadrature, and are mainly proportional to the imaginary part of the electro-optic coefficient $C_{\text{ind.}}^E$, as given by

$$P_z^{\text{trans.}} = i\, n_R^{-2}\, (C_{\text{ind.}}^E)''\, E_x^L (E_x^S)^* \quad (7)$$

Consequently, we assume that this transverse part of these longitudinal polarisation waves should be the origin of the coherent microwave radiation, observed in our SRS experiments with weakly compressed hydrogen.

Besides, these two progressive polarisation waves should have theoretically an infinite wavelength, but on the physical point of view, this wavelength must be finite and very large compared to the light wavelengths. This is obtained by taking into account of an anomalous relativistic Doppler effect in a dielectric medium, as proposed in ref. [9]. Then, the angular frequency $\omega_o$ of the microwave emission is related to the frequency $\omega_R$ by the following equation

$$\omega_o = (1 - \beta^2)^{1/2}\, \omega_R \quad (8)$$

where $\beta$ is the ratio between the phase velocity $V_o$ of the polarisation waves and c; this parameter must be very close from the unity value in our model, what is effectively obtained in the relation given below

$$1-\beta = 1/2\ (\omega_o/\omega_R)^2 \approx 1/2\ (20\ 10^9/\ 4155\ c)^2 \approx 10^{-8}\text{-}10^{-7}$$

Furthermore, the microwave radiation, which is assumed to be of Cherenkov-type, should be emitted inside the Cherenkov cone, simultaneously with the SRS process, at least to all appearances, as predicted by Ginzburg et al. in their theory [14-15]; our experimental results, shown in Fig.6, seem to agree with their assumption.

Finally, concerning the exponential decreasing of the microwave signal beyond the pressure of 8-10 bar, it is possible to explain this behaviour by assuming that the quasiparticle density is strongly reduced by the increasing effects of dephasing and non-radiative relaxation of the vibrational energy, due to the Van der Waals forces. As a matter of fact, in the pressure range of 3-10 bar the collisional narrowing is the main effect [16], and the intensities of the stimulated Raman waves and of the microwave emission, are both strongly increasing. However, around the pressure of 8-10 bar the influence of the Van der Waals forces becomes sensitive, because the value of the collision rate $\Omega$ is equal to about the reciprocal of the average of the quasiparticle lifetime $\tau_R$, which corresponds to the relation $\Omega\tau_R \approx 1$; but at higher pressure this equality must be replaced by the inequality $\Omega\tau_R > 1$, which shows that the Van der Waals forces are preponderating.

**IV-Discussion**.-First of all, it seems to us that our model might be drawn to the one developed by Dumartin, who had attempted to interpret his experimental results about the process of Stokes absorption, observed in several liquids during experiments of SRS associated with a broad spectral continuum, when the laser intensity is slightly above the SRS threshold [17-18]. For instance, Fig.7 displays the spectrum of the Stokes absorption band of liquid benzene, the bandwidth of which is one order of magnitude larger than the width of the Stokes emission line in spontaneous Raman scattering. To explain this process, Dumartin had shown that near the SRS threshold, the population of vibrationally excited molecules might be sufficiently high during very short times ($10^{-12}$-$10^{-11}$ s), compared to the duration of the laser pulse, to assume the existence of a "population inversion", giving rise to very broad Stokes absorption bands [18].

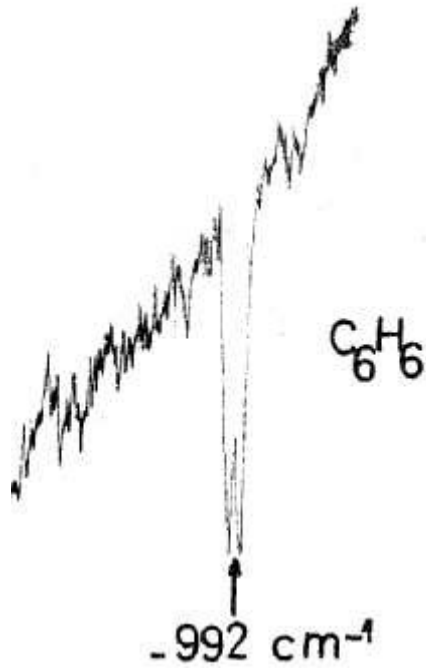

Fig.7- Microdensitometer of the stimulated Stokes absorption band, obtainted in liquid benzene [18]; note the weak stimulated emission inside the strong absorption band.

On the other hand, a comparison might be made between our results about a coherent microwave emission by SRS in hydrogen , and observations of forward conical emissions in barium or sodium dense vapours illuminated by dye lasers at a slightly shorter wavelength than those of the resonance transitions; these emissions are picked at a longer wavelength, with same shifts from the resonance transitions, as those used for their excitation [19-22]. As a matter of fact, these last phenomena have been interpreted as Cherenkov-type processes, the origin of the conical emissions being at the surface of self-focusing filaments [20-22]. However, in our case we propose also to take into account of a Cherenkov-type process, but the microwave radiation must be emitted in volume and inside the Cherenkov cone (see Fig.5), without self-focusing and filaments in hydrogen. Moreover, Golub et al. have calculated the refractive indices at different detunings to the red of the resonance transition, and have found that these parameters show anomalous dispersion profiles; they have also calculated the half- angles of the Cherenkov cones, given by the ratio $n_L/n_C$ where $n_C$ is the maximum of the anomalous profile, and have observed a good agreement with the experimental values [21]. On the other hand, in our case we assume that the half- angle of the Cherenkov cone is given by the ratio $1/n_R$, since the process of microwave emission takes place exactly at the resonance frequency $\omega_R$.

Besides, we think that our model might be drawn somewhat to the interpretation proposed by Le Quéau et al. for explaining observations of a coherent auroral kilometric radiation [23-24]. As a matter of fact, these authors have proposed a model based on the maser synchrotron instability, as follows: they assume that electron beams arising from the solar wind are weakly accelerated to several keV by an electric field, in a region of the magnetosphere where this field is parallel to the terrestrial magnetic field force lines, and have a spiral motion around the magnetic field lines. Therefore, if the electron distribution function has a positive slope between two values of the perpendicular electron velocity, which should correspond to a "population inversion" having a short lifetime of about 1 ms, an energy exchange between particle and wave is possible, showing a peak when the following resonance condition is fulfilled:

$$\omega_X = (1 - V^2/c^2)^{1/2} \omega_c$$

where V is the total electron velocity, $\omega_X$ the angular frequency of the coherent kilometric radiation and $\omega_c$ the cyclotron frequency of electrons.

As can be seen, there are several similarities between this last model and our own; however, one of the main differences is due to the fact that in the former the Doppler effect is weakly relativistic, while in our model it is necessary to take into account of an ultrarelativistic Doppler effect.

Finally, let us note that recently it has been possible to obtain coherent radiations of GW power in the far ultraviolet region, owing to a high-gain free electron laser working in a single pass [25-27]. In this device the linear electron beam, emitted as short bunches of ps duration, is accelerated to a velocity V', very close to c, and forced to oscillate in the undulator made up of a periodic sequence of alternating transverse magnetic fields. Then, the resonance wavelength $\lambda_{ph}$ of the light emission is proportional to the quantity $(\lambda_u/2)[1-(V'/c)^2]$, where $\lambda_u$ is the period length between the permanent magnets of the undulator. For instance, in the last experiments, radiation at 32 nm wavelength has been observed, the light pulses having a duration of about 25 fs, while the electron bunch contained a 50 fs long leading spike [27].

Moreover, it is interesting to note that the "normalized emittance" of this free electron laser is proportional to the relativistic factor $(V'/c)[1-(V'/c)^2]^{-1/2}$, as given in ref.[27]; we feel that this parameter might be somewhat equivalent to the one used below for explaining the physical meaning of the third term in the relation (9), once multiplied by the relativistic factor $\beta^2(1-\beta^2)^{-1/2}$.

**V-Remarks about the SRS gain.-**It is well known, when there is a jump at the SRS threshold in a dielectric medium, that the experimental values of the Raman gain for the Stokes wave are very much higher than those calculated by means of the classical SRS theories [28]. These disagreements were explained by many authors, by taking onto account only the process of self-focusing due to the variation of the refractive index at high laser intensity, giving rise to a strong increasing of the laser electric field [29].

Nevertheless, it seems to us that an additional contribution might be also taken into account, which should be the origin of an enhancement of the Raman gain. As a matter of fact, we can calculate the polarisation density at the Stokes frequency, given by the relation (1c), by taking into account the relations (1a), (5) and (6). For obtaining the total polarisation density, we must add also to this computation, the imaginary part of the third order susceptibility at the Raman frequency (fourth term in the relation (9)), which is the origin of the exponential gain for the Stokes wave in the classical SRS theories, based only on the approximation of the steady state. Then, the following relation is obtained:

$$(P_x^S)^* = [(n_L^2 - 1)/4\pi](E_x^S)^* - 8\pi n_R^{-2} |C_{ind.}^E|^2 |E_x^L|^2 (E_x^S)^*$$
$$\mp i\, 32\, \pi^{5/2}\, n_R^{-5}\, B^3\, \omega_R^{-3} |C_{ind.}^E|^2 |E_x^L|^2 (E_x^S)^* + (\chi_{Ram.}^{(3)})'' |E_x^L|^2 (E_x^S)^* \quad (9)$$

In this expression the first term denotes the linear part of the Stokes polarisation, which is the source of spontaneous Raman scattering. The second term is a nonlinear part, which depends on the square of the induced electro-optic coefficient, and might be an additional contribution to the amplification of the Stokes wave, effective only for very short times, e.g. during parametric instabilities of about $10^{-13}$-$10^{-12}$ s. Finally, the third term is a in phase quadrature part of the polarisation, and should be a longitudinal component corresponding, once multiplied by the relativistic factor $\beta^2(1-\beta^2)^{-1/2}$, to the translation energy of the whole medium, which must cancel during instability durations, the momentum associated to the progressive quasiparticle waves, according to the theoretical discussions submitted by de Broglie about some relativistic formulas [30].

**Conclusion.-**In this paper experimental results are presented, concerning observations for the first time, to our knowledge, of a coherent microwave radiation during SRS experiments in weakly compressed hydrogen, concomitant with the emission of stimulated Raman lines. The intensity of this microwave radiation had been studied as a function of the gas pressure, the laser and Stokes beam powers, and the distance between detector and propagation axis of the laser beam. Moreover, an estimate of the wavelength had been obtained, the value of which is of the centimetric order of magnitude; an estimate of the peak power of this microwave emission had been also determined to be in the order of magnitude of 0.1-1 mW.

An attempt to interpret the origin of this coherent microwave is proposed, on the basis of a phenomenological model, taking into account of an intermediate quasiparticle state, due to the influence of the vibrational anharmonicity of hydrogen molecules, when the density of vibrationally excited oscillators is high. Besides, we assume that the duration of this intermediate state should be very short ($10^{-13}$-$10^{-12}$ s).

On the other hand, it is stated that the electro- mechanical action of the laser light on the quasiparticle system gives rise to a "piezoelectric" character for this system; in addition, a nonlinear coupling of anharmonic type between the Stokes wave and the two progressive waves associated to the quasiparticle system is taken into account, together with the existence of a macroscopic longitudinal electric field. Then, the quasiparticle system might be called an "anharmonic longitudinal polariton" system. A Cherenkov-type effect and an anomalous relativistic Doppler effect are also taken into account.

Therefore, we assume that the source of the coherent microwave radiation is the in phase quadrature part of the longitudinal component of the polarisation waves, associated to the quasiparticle system . On the other hand, one can note that according to our model the microwave radiation should be emitted during very short times, like parametric instabilities which should appear as a continuous series, during the time of propagation of the exciting laser beam in the dielectric medium.

Concerning the SRS processes, recall that above the SRS threshold anomalous large gains had been observed, which had been explained for a long time, only by the influence of self-focusing processes. However, it seems to us that it is necessary also to take into account of an additional contribution arising from the influence of the nonlinear coupling cited above, together with a cooperative effect.

Finally, let us mention that in the case of spontaneous Raman scattering with crystals, observations of new Raman lines over the normal number of lines related to crystal symmetries, and of enhanced Raman scattering, have been correctly interpreted for the first time by Poulet [31]. This author had shown that in the case of piezoelectric cubic solids, these two effects can be accounted for, in the case of the first one by the action of a longitudinal electric field induced by the exciting light beam, and of the second by the influence of the whole medium due to a linear electro-optic coefficient, related to this longitudinal electric field [31]. In other respects, Burstein et al. had been able to interpret experimental results in the semiconductor GaAs, on the basis of the $k \approx 0$ coupled plasmon-(LO) longitudinal-optical–phonon collective modes [32]. Furthermore, Birman et al. had propounded a theoretical explanation for the effect of enhanced Raman scattering in crystals, by using the exciton representation as intermediate state; they have shown that "for longitudinal polar optic vibration, there is an additional contribution to the Raman tensor due to the associated electric field" [33].

Bibliography

-----------

[1]- B. Oksengorn, C.R. Acad. Sci. Paris, 320 Série II b (1995) 509-514

[2]- G. Bret, Annales de Radioélectricité, T. XXII (1967) 236-288

[3]- A.A. Manenkov, V.A. Milyaev, G.N. Mikhailova, and S.P. Smolin, Soviet Physics, JETP Lett.,16 (1972) 322-325

[4]- L.V. Keldysh, A.A. Manenkov, V.A. Milyaev, and G.N. Mikhailova, Soviet Physics, JETP, 39 (1974) 1072-1078

[5]- J.O. Hirschfelder, C.F. Curtiss, and R.B. Bird, Molecular Theory of Gases and Liquids ; John Wiley § Sons, Inc. : New-York, 1964

[6]- J.C. Kimball, C.Y. Fong, and Y.R. Shen, Phys. Rev. B, 23 (1981) 4946-4959

[7]- N.L.Boling, A.J. Glass, and A. Owyoung, IEEE J. of Quant. Electr., QE 14 (1978) 601-608

[8]- P.A. Franken and J.F. Ward, Rev. of Modern Phys., 35 (1963) 23-39

[9]- B. Oksengorn, Chem. Phys. Lett., 1 (1968) 591-593

[10]- J. Wenk and L. Merten, Phys. Stat. of Sol. (b), 93 (1979) 175-181[11]- V. Schulz and L. Merten, Phys. Stat. of Sol. (b), 115 (1983) 225-233

[12]- L. Merten and R. Kotzott, Phys. Stat. of Sol. (b), 120 (1983) 481-489

[13]- L. Merten and W. Goltsche, Phys. Stat. of Sol. (b),121 (1984) 471-479

[14]- V.L. Ginzburg and V.Ya. Eïdman, Soviet Physics, JETP, 36 (1959) 1300-1307

[15]- V.L. Ginzburg, Waynflete Lectures on Physics, Vol. 106; Pergamon Press Ltd, Oxford, 1983

[16]- P. Lallemand, P.Simova, and G. Bret, Phys. Rev. Lett., 17 (1966) 1239-1241

[17]- S.Dumartin, B. Oksengorn, and B. Vodar, C.R. Acad. Sci. Paris, 261 (1965) 3767-3770

[18]- S. Dumartin, Thèse d'Etat, Université Paris 11, Orsay, 1972 (unpublished)

[19]- C.H. Skinner and P.D. Kleiber, Phys. Rev. A, 21 (1980) 151-156.

[20]- I. Golub, R. Shuker, and G. Erez, Opt. Commun., 57 (1986) 143-145

[21]- I. Golub, G. Erez, and R. Shuker, J. of Phys. B,19 (1986) L115-L120

[22]- V.I. Vaichaitis, M.V. Ignatavichyus, V.A. Kudryashov, and Yu.N. Pimanov, Soviet Physics, JETP Lett., 45 (1987) 414-417

[23]- D. Le Quéau, R. Pellat, and A. Roux, Phys. of Fluids, 27 (1984) 247-265

[24]- D. Le Quéau, R. Pellat, and A. Roux, J. of Geophys. Res., 89 (1984) 2831-2841

[25]- V. Ayvazyan et al., Phys. Rev. Lett., 88 (2002) 104802,1- 4

[26]- V. Ayvazyan et al., Eur. Phys. J., D20 (2002) 149-156

[27]- V. Ayvazyan et al., Eur. Phys. J., D37 (2006) 297-303

[28]- Y.R. Shen, The Principles of Nonlinear Optics ; John Wiley § Sons, Inc.: New-York, 1984

[29]- Y.R. Shen, Prog. Quant. Electr., 4 (1975) 1-34

[30]- L. de Broglie, C.R. Acad. Sci. Paris, 264 Série B (1967) 1173-1175. - 265 Série B (1967) 437-439. -  265 Série B (1967) 589-591

[31]-  H. Poulet, Annales de Physique, 10 (1955) 908-967

[32]-  E. Burstein, A. Pinczuk and S. Iwasa, Phys. Rev., 157 (1967) 611-614

[33]-  J.L. Birman, and A.K. Ganguly, Phys. Rev. Lett., 17 (1966) 647-649